# A Spreadsheet Auditing Tool Evaluated in an Industrial Context

*Markus Clermont, Christian Hanin, Roland Mittermeir*
*Universität Klagenfurt*
*Universitätsstraße 65- 67*
*A-9020 Klagenfurt*
*Austria*
[mark@isys.uni-klu.ac.at](mark@isys.uni-klu.ac.at)

**ABSTRACT**

*Amongst the large number of write-and-throw-away-spreadsheets developed for one-time use there is a rather neglected proportion of spreadsheets that are huge, periodically used, and submitted to regular update-cycles like any conventionally evolving valuable legacy application software. However, due to the very nature of spreadsheets, their evolution is particularly tricky and therefore error-prone.*
*In our strive to develop tools and methodologies to improve spreadsheet quality, we analysed consolidation spreadsheets of an internationally operating company for the errors they contain. The paper presents the results of the field audit, involving 78 spreadsheets with 60,446 non-empty cells. As a by-product, the study performed was also to validate our analysis tools in an industrial context.*
*The evaluated auditing tool offers the auditor a new view on the formula structure of the spreadsheet by grouping similar formulas into equivalence classes. Our auditing approach defines three similarity criteria between formulae, namely copy, logical and structural equivalence. To improve the visualization of large spreadsheets, equivalences and data dependencies are displayed in separated windows that are interlinked with the spreadsheet. The auditing approach helps to find irregularities in the geometrical pattern of similar formulas.*

## 1 INTRODUCTION

Spreadsheets are a main factor contributing to the success of the personal computers. Today, they might be considered to be the most successful end-user programming tool. Each year, millions of spreadsheets are developed. Lots of them are small and used for one-time calculations, but there is a substantial number of spreadsheets that are large and complex.

These are usually strategically important and contain both large and complex calculations. These sheets might also be quite long-lived. Hence, undergo similar evolutionary steps as conventional software. In [Tampoe, 1996], spreadsheets are presented as strategic management information systems. Thus, erroneous spreadsheets, notably those long-living ones will have severe consequences.

The strategic spreadsheets we analysed generally consist of two parts: The first part is very large, but relatively uniform. It serves to gather data and to perform some quite simple calculations. This part can be spread out on very large areas of a sheet. It needs not to be contiguous, but it tends to be so. In the sheets we analysed, up to 20 columns and more than 200 rows are common in this part. The second part is much smaller, but contains more complex calculations. Examples are calculation of enterprise-specific financial ratios, time-series analysis or the generation of check-sums. While the first part confronts the auditor with a complexity 'of size', the complexity of the second part is due to a limited number of complex calculations.

Of course, one must not over-generalize from the sample of 78 sheets we analysed over a period of three months. But it seems fair to assume that any developer of a sheet that is repeatedly used strives to for an arrangement that is somehow related to the semantics of the sheet. Normally, this arrangement follows a well-understood business pattern. With large sheets, such business logic leads to arrangements where data-entry cells, cells immediately dependent on these data entries used for preparatory operations, and cells performing the final modelling or analysis are allotted to distinct, well identifiable locations (to avoid confusion we avoid the term "area" at this moment) or laid out in a regular pattern.

Moreover, the sheets we analysed seem to be typical for sheets involved in financial or commercial applications. In [Filby, 1993] numerous applications of spreadsheets in science and engineering are presented. These spreadsheets are used in physics, chemistry and other sciences, because they are a more usable alternative to FORTRAN-programs and because they incorporate already the (graphical) representation of their result. As these spreadsheets specialize on complex calculations, we do only find the second part mentioned above. The data-entry portion is comparatively simple in these cases.

Our auditing methodology reduces the complexity of size by banking on regularities in the cell content. Similar cells are grouped into so-called logical equivalence classes. Cells that are in the same logical equivalence class are presented to the user by a single abstract unit, the logical area. When the logical areas are highlighted on the spreadsheet, the user can easily spot inconsistencies between the geometrical pattern of formula usage and the conceptual model they had in mind. Complex calculations that occur only in a few cells of the spreadsheet still have to be examined on a cell-by-cell level (c.f. [Panko, 1997]).

The rest of the paper is organized as follows: Section 2 points out the main sources of errors discovered in our field audit. In section 3 we briefly explain our auditing technique and present the toolkit used. Additionally we describe the reviewed spreadsheets and the context of their use. In section 4 the results of the field audit are presented and we try to categorize the revealed errors. Section 5 addresses the methodological issues involved with the experiment.

## 2 ERROR SOURCES

Indirectly, the ease of creating spreadsheet programs is the most important source of errors: Spreadsheet programs can be created without a great deal of IT-training and even complex models can be implemented by rather simple means.

The low level of the spreadsheet users IT-training will make them neglect important tasks like analysis, documentation and in-depth testing, as it seems that there is no direct relation between these tasks and the success of a spreadsheet program. [Nardi, 1990] states, that the spreadsheet is also an important modelling tool for the users. Thus, the spreadsheet program is quite often all in one: the modelling tool, the design and the implementation of an information system.

This procedure is in sharp contrast to the importance of spreadsheets for organizations. [Gable, 1991] analysed the importance of 400 spreadsheets for their organizations, and came to the conclusion, that more than 50% of them were considered to be very important. [Chan, 1996] interviewed more than 200 spreadsheet users on their estimation of the cost of an error in their spreadsheet. 4.6% estimated the potential damage is more than 1,000,000 USD. In [Panko, 2002] some drastic examples for spreadsheet errors that economically damaged the affected organization, are reported.

### 2.1 Complexity

Although spreadsheets are not very complex to create, the mechanism of absolute and relative cell references will rapidly lead to a high degree of complexity within them. Spreadsheet users are generally not aware of that fact. Thus, mistakes, that have been made anywhere in the underlying model, will be propagated.

The principle of locality, an important concept for reducing the complexity of software, is not part of the spreadsheet model, i.e. any other cell anywhere on the spreadsheet can freely access the result value of a certain cell. Hence, the effects of an error in an arbitrary cell will potentially influence one or more results of the spreadsheet irrespective of their "distance" to the erroneous cell. Worse, the effect of an error might show at a different place than the error itself, thus further increasing the complexity of identifying faults.

There are techniques to reduce the complexity of spreadsheet programs, by forcing the spreadsheet user to build modular spreadsheets (see e.g. [Knight, 2000], [Janvrin, 2000], [Stadelmann, 1993], [Wilde, 1993]). However, these techniques are not widely used yet. In contrast to these techniques, we do not aim to change spreadsheet users. We suggest taking the sheets they developed on an as-is basis. We do assume, however, that even computing-laypersons do not spread out their calculations on the sheet in a random order. In contrast, we assume that they use the (two) dimensions of the sheet in an intelligent manner to floor plan the layout of their calculations.

## 2.2 Copy and Paste

Usually spreadsheets are created by defining a formula and then copying this formula into the cells were the same or a similar functionality is expected. The same formula tends to occur very often, but the geometric distances between these occurrences can be quite large.

Thus, the copy/paste mechanism is somehow similar to the use of subroutines (rather macros) in conventional software. However, there are some important differences that entail dangerous side effects:

- If the copied formula is erroneous, the error is replicated, too.
- Past the copy operation, the duplicated cells forget from where they originated.
- If an error is detected and corrected only at one place, all the other copies of this formula remain still erroneous.
- Error corrections might be done on the value level only, thus leading to incorrect sheets in future instantiations.

## 2.3 Error Correction

As in conventional software, we identified error correction as an important source of future errors in spreadsheets. Spreadsheet users tend to check their spreadsheets on the numerical level. When mismatches between their expectations and the shown result occur, they often fail to debug the formula. This might be considered to be too time consuming, because the real cause of the wrong value shown at the given cell is not obvious. Therefore, they just overwrite a formula with a constant value. As a consequence, the error is currently corrected and the current sheet shows correct computations. However, further changes to the spreadsheet will not be reflected in this cell. Thus, a new, latent error is introduced.

Again, we do not want to over-generalise. However, considering the training of the clerks working with these sheets, it is no wonder that they focus on the value domain of their sheets. Considering the value domain, we have to credit them with respect for highest diligence and care. Being no programmers though, they did not see that below this value domain there is a model domain (or "program domain") expressed by the network of formulas tightly interwoven by linkages of references and data-flow. Therefore, the problem that their models are correct only, if these models are correct on the model (or program-) domain first, was something they have only gradually accepted during the time they worked with us.

## 2.4 Maintenance

A given long-living spreadsheet usually continues to evolve. As we already learned from conventional software [Parnas, 1994], software ages with maintenance. In order to keep up with evolving requirements, ongoing adjustments must take place. Changes in the environment of spreadsheet programs, like new tax-rates or new organizational structures, will force the spreadsheet users to maintain the spreadsheet.

However, the lack of documentation makes it hard for spreadsheet authors to understand the effect of changing a single cell has on the rest of the spreadsheet. If the maintainer is not the original author, these problems are further aggravated. Maintainers do not know about the authors' conceptional model of the spreadsheet. Thus, they have to perform maintenance based on their assumptions. It is obvious that this procedure will blur the initial spreadsheet model and makes it 'age', as it is stated by [Parnas, 1994] for conventional software, quite rapidly.

Another common maintenance operation is the intended change of the functionality of a certain spreadsheet program, in order to make it applicable for problems that are similar to the original problem. Therefore, only those parts of the spreadsheet are modified, where changes are obviously needed. Other parts are not modified, which can entail misunderstandings and errors in further maintenance cycles.

Obviously, the actual spreadsheet development process does not support the high importance of spreadsheet programs. A methodical approach, thorough testing and sufficient documentation, steps common for raising the quality of conventional software, are hardly ever used in spreadsheet development. The short maintenance cycles and the lack of modularisation also promote the introduction and propagation of errors.

## 3 ORGANIZATIONAL ENVIRONMENT OF THE FIELD AUDIT

This section will introduce the auditing technique, the organizational environment of the audit, and the characteristics of the audited spreadsheets.

### 3.1 Auditing Technique

As already mentioned in section 2.4, misunderstandings regarding the spreadsheet model will make spreadsheet maintenance error prone. Further, testing of spreadsheets is complicated, as the internal logic is not clear to the tester. We developed an auditing technique to reveal the spreadsheet model by showing the occurrences of similar formulas throughout the spreadsheet. Thus, regular patterns, or irregularities can be spotted at first sight.

Irregularities generally do not indicate an error, but they indicate a dangerous spot that has to be checked, whereas regular patterns are a hint for a direct manifestation of a conceptual model on the spreadsheet. As effective auditing of spreadsheets is stated to be an expensive and time consuming task [Panko, 1997], our auditing technique will reduce the number of cells to be examined by finding the potentially dangerous areas and focussing the auditors' attention on these areas. Further, we offer another view on the conceptual model. It shows the data-flow, i.e. the dependencies, between these regular areas.

By understanding the abstract representation our tool provides, the auditor can comprehend the architecture of the spreadsheet. Thus, error correction and maintenance are supported, as the maintainer is aware of regular patterns of formula-occurrences. This helps in comprehending sheets originally written by others.

Symptoms of errors are often erroneously corrected by overwriting the correct formula with a constant value or another formula. In these cases, the problem is only aggravated, because the formula just showing an incorrect value due to an error in another cell is destroyed by this pseudo-corrective act in the value domain. As auditing spreadsheets by finding irregularities is not based on symptoms, but on causes of errors, correction can be focussed and is thus easier to perform.

Our technique identifies regular structures in the spreadsheet. These regular structures, so called logical equivalence classes, are sets of similar cells. These similar cells do not have to be neighbours, but we noticed that on large sheets

- They are either neighbours on the layout, or
- They are distributed in a regular pattern, or
- Their occurrence is limited to a certain area of the spreadsheet

Of course, none of these points need to be the case. But for the majority of logical equivalence classes at least one of these properties applies.

Above, we defined the logical equivalence class to be a set of similar cells. The similarity is defined by comparing the formulas. We consider the following three kinds of equivalence classes:

1. **Copy-Equivalence** exists, if the formulas are absolutely identical (i.e. the cell contents has been copied from one cell into the other, either by copy and paste, or by retyping the same formula).
2. **Logical- Equivalence** exists, if the formulas differ only in constant values and absolute references
3. **Structural- Equivalence** exists, if the formulas consist of the same operators in the same order, but the operators may be applied to different arguments.

By comparing the partition of cells into logical equivalence classes with their geometric distribution on the spreadsheet, inconsistencies can be easily spotted. E.g. if a set of cells in a column is copy-equivalent, but there is one cell interspersed that contains a different formula or a constant, this indicates an inconsistency that has to be further investigated.

### 3.2 The Toolkit

In order to support the auditing process we developed a toolkit that automatically performs the partitioning into equivalence classes. The toolkit consists of three main parts: A structure browser (see Figure 1) to show the decomposition of the spreadsheet into equivalence classes, a dependency viewer that displays the data flow graph between these dependencies, and the spreadsheet itself giving feedback to the auditor by highlighting the cells that are in the equivalence class that is currently selected in the structure browser.

The structure browser uses the equivalence class hierarchy (see Figure 2) to give a hierarchic view. As the auditors are able to expand and collapse the nodes in the structure browser they can zoom into certain equivalence classes, whilst viewing the remaining nodes on a higher level of abstraction. Only those nodes that are visible in the structure browser are displayed in the dependency viewer.

As we used only an α-Version of our tool for the audit, the technical skills of the auditor were highly needed. The integration between the dependency viewer and the structure browser was rather rudimentary, by generating files in the structure browser and displaying them with the free graph layout software Dotty (see [Ganser, 1999]). In the subsequent versions of our auditing tool we aim for a tighter integration between dependency viewer, structure browser and spreadsheet.

### 3.3 Organizational Environment

Auditing was performed from April until August 2001 by a computer-science student in the sixth semester. The auditor was assigned to the accounting department of an international cooperation with headquarters in Vienna where he could work desk-to-desk with the spreadsheet producers. The contact with the tool developers was by e-mail and by regular visits. He examined three voluminous Excel-workbooks (see section 3.4) that are mainly used for consolidation. The three workbooks consisted of 78 worksheets, with 60,446 non-empty cells.

The identified errors were coarsely categorized by their immediate impact into qualitative and quantitative errors (see [Teo, 2000]), and by their origin into the following categories (see [Ayalew, 2000]):

- Constant instead of formula
- Constant instead of reference
- Reference to empty cell
- Formula copied too far
- Other

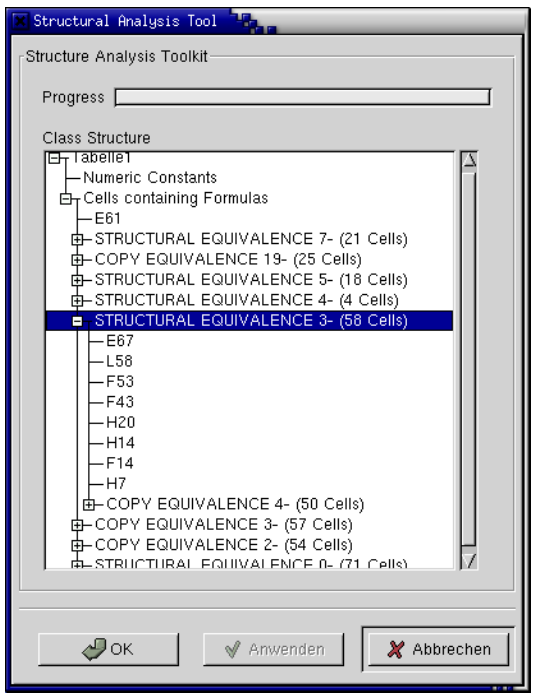

Figure 1: The Structure Browser with example-data

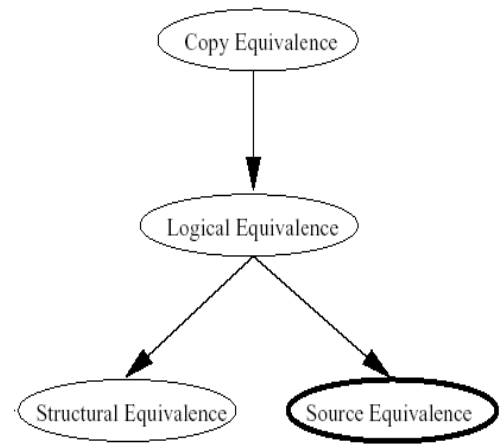

Figure 2: Relevant portion of the partial order between logical equivalence classes.

The consequence of a quantitative error is an erroneous result of a cell on the spreadsheet, i.e. a wrong result in the value domain. This does not necessarily relate to an error within the cell that contains the quantitatively erroneous formula. As already mentioned above, the error and the symptom of the error can turn up in different cells.

In contrast, qualitative errors will not immediately entail a wrong result in the value of any cell. However, they are (potential) errors in the model. When maintenance is performed, these qualitative errors usually turn into quantitative errors, i.e. somewhere on the spreadsheet a corrupted value will be displayed. An example for a common qualitative error is an erroneous expression in one branch of an if-statement in a certain cell. As long as the erroneous branch is not activated, there is no symptom of fault for this cell.

### 3.4 Auditing Process

Before the audit started, the auditor, who had only little bookkeeping experience, discussed the basic idea and functionality of each workbook with the respective author. Additionally, the author was interviewed about the lifespan of the workbook, the usual maintenance cycle and the number of users.

Then, for each spreadsheet in the workbook, the following characteristics were documented: Dimension, Number of occupied cells, Number of formulas, constants and literals. At first, the correctness of the displayed values was checked. Special attention was put on wrong sums, wrong formatting and errors that were reported by Excel.

After these routine checks in the value domain, the toolkit described in section 3.2 was applied. The so discovered irregularities where then discussed with the spreadsheet authors, to find out, if the detected irregularities were deliberately introduced or whether they have to be corrected and counted in the error statistics.

Thus, the auditor had a lot of discussion with the domain specialists who created the spreadsheets. No error was documented that was not verified by the spreadsheet creator. The identified errors were collected in an error database. For each error we gathered information about the location, the kind of error, and its impact. Additionally, a short description was also stored.

As an error can be multiplied by copy and paste operations, we distinguish between errors and error classes. Copy-equivalent erroneous formulas are counted as one error-class. Hence, the error-class corresponds to the unique source of an error that can be copied into several cells. The term error is used to count each of the error-instances within the respective error class. Thus, each error represents an erroneous cell.

### 3.5 Examined Spreadsheets

The audit examined three large excel workbooks. Each of them was used to gather data from various departments of the company and to calculate different financial ratios at the corporate level. These financial ratios are an important base for strategic decisions. The workbooks analysed served the following purpose:

- **RAT-2001** calculates a financial statement. Data is aggregated from sub-sheets that correspond to the enterprise's organization. Hence, there are worksheets for different business-units (BU) and corporate sectors. These worksheets are aggregated to calculate the financial statement of each division. The spreadsheet has been in use for one year so far. There is extensive maintenance each month. The company's annual budget processed by these spreadsheets is about €150,000,000.

- **TP-Report** was in use for three months when we examined it. The lifespan of the spreadsheet was considered to be unlimited. When audited, the author was the only user. But it was planned to delegate maintenance of particular worksheets to other employees. The sheet accumulates data from four other workbooks that are maintained by four different persons. During our study the workbook has been fundamentally changed, so we re-audited it. In the results we only mention the latest version audited.

- **AB-Market** performs material costs analysis. It is in use since 1999 and modified each year, before budgeting is done. A copy of the workbook is sent to each branch office where its input cells are filled in by at most three employees. The completed/updated workbooks are sent to the author again, who merges the copies into a single workbook. The data obtained by this procedure is used to analyse cost of raw material of the various factories. For the analysis, additional information, such as current and forecasted volume, costs, price per unit, and average prices are added to the workbook. This information is extracted from the companies SAP-based information system. The workbook calculates a budget target for each factory that can be compared to the planned budget. The calculated budgets' values are about €13,000,000 each.

## 4 RESULTS

Concerning error statistics, the results we obtained correspond to the findings of earlier studies and the reports of practitioners (see [Panko, 2002], [Butler, 2000]). The overall error rate was 3.03% of the non-empty cells. However, we did not find any tremendous erroneous result values that might have had severe negative effects on the company. What we found though was a very high number of qualitative errors with the potential to become quantitative errors in the next (or future) maintenance cycle(s).

Thus, the numerical test that each workbook undergoes after each round of modifications becomes more difficult, and, as we argued above, the increasing number of "corrected" errors tends to introduce more qualitative errors in the model (see section 3.3). This vicious circle cannot be interrupted without corrections of the spreadsheet model.

### 4.1 Overview of Results

In 78 audited spreadsheets 109 error classes with 1832 occurrences were identified (see table 1). As the workbooks have usually consisted of similar spreadsheets, the occurrence of one error class is not limited to one spreadsheet. We identified several error classes that were copied into different spreadsheets of the same workbook.

The workbook TP-Report was still under construction when our study finished and so many of the identified problems were immediately corrected. This explains, so many error classes were detected in this workbook. The workbook AB-Market has been re-designed a short time before our audit took place. Hence, there was only a small amount of errors in the model.

The distribution of errors in the audited workbooks is given in absolute numbers in Table 1, whereas ´Table 2 gives the relative distribution with percent-values.

| Workbook | #Cells | #Occupied | #Formula | #Literals | #CE | #Error Classes | #Errors |
|---|---|---|---|---|---|---|---|
| RAT-2001 | 56,485 | 19,444 | 12,382 | 7,062 | 814 | 21 | 257 |
| TP-Report | 69,835 | 23,502 | 16,873 | 6,629 | 950 | 83 | 1,561 |
| AB-Market | 66,385 | 17,500 | 7,174 | 10,326 | 95 | 5 | 14 |
| *Total* | *192,705* | *60,446* | *36,429* | *24,017* | *1,859* | *109* | *1,832* |

Table 1: Error Distribution, absolute

By classifying the errors and error classes into quantitative and qualitative errors, we obtained the distribution given in Table 3. The classification into the error-categories listed in section 3.3 is given in Table 4. The category *Others* consists of a wide diversity of error classes with patterns more or less unique for the individual instances.

| Workbook | #Cells | #Occ. | #Formula | #Literals | CE/Formula | #Error Classes | #Errors |
|---|---|---|---|---|---|---|---|
| RAT-2001 | 56,485 | 34% | 64% | 36% | 6.6% | 21 | 1,3% |
| TP-Report | 69,835 | 34% | 72% | 28% | 5.6% | 83 | 6,7% |
| AB-Market | 66,385 | 26% | 41% | 59% | 1.3% | 5 | 0,08% |
| *Total* | *192,705* | *31.37%* | *60.27%* | *39.73%* | *5.1%* | *109* | *3.03%* |

Table 2: Error Distribution, relative (#Errors is given relative to occupied cells)

| Workbook | Category | Error Classes | Errors |
|---|---|---|---|
| RAT-2001 | Qualitative | 7 | 84 |
| | Quantitative | 14 | 183 |
| TP-Report | Qualitative | 73 | 1503 |
| | Quantitative | 10 | 58 |
| AB-Market | Qualitatve | 5 | 14 |
| | Quantitative | 0 | 0 |
| *Total* | *Qualitative* | *85* | *1591* |
| | *Quantitative* | *24* | *241* |

Table 3: Error classification into qualitative and quantitative errors

In order to check the effectiveness of the auditing technique, we calculated the Copy-Equivalence to Formula ratio, i.e. the average size of each copy equivalence class. In the average, each copy-equivalence class contains 5.1 formulas. Thus, only every fifth formula cell of the spreadsheet had to be checked in detail. Of course, this measure is blurred, as there are certain formulas, e.g. check-sums or other validation formulas, that occur only once, whilst others occur more than 20 times. For multiple occurrences of the same formula it had only to be checked, if they are used in the right place. The frequency of occurrence of error classes relative to copy-equivalent classes is obviously correlated to the frequency of errors relative to formulas. This seems to support our assumption that errors are likely to be multiplied by copy & paste. However, as it is shown by Table 5, the workbook *AB-Market* does not follow this trend. We argue that this is because of the 'youth' of this workbook. The errors detected seem to be mainly in checksums and thus, not copied over many cells.

| Error Category | Error Classes | Errors |
|---|---|---|
| Constant instead of formula | 16 | 1222 |
| Constant instead of reference | 8 | 78 |
| Reference to empty cell | 8 | 78 |
| Formula copied to far | 24 | 215 |
| Other | 53 | 239 |

Table 4: Error distribution by error category

| Workbook | #Formula | #CE | #Error Classes | CE/Formula | Error Classes / CE | Errors / Formula |
|---|---|---|---|---|---|---|
| RAT-2001 | 12382 | 811 | 21 | 6.6% | 2,6% | 2,07% |
| TP-Report | 16873 | 950 | 83 | 5.6% | 8,7% | 9,25% |
| AB-Market | 7174 | 95 | 5 | 1.3% | 5,2% | 0,19% |
| *Total* | *36429* | *1859* | *109* | *5.1%* | *5,9%* | *5,02%* |

Table 5: Error Class Distribution, relative to copy-equivalence classes

## 5 TOOL ASSESSMENT

In spite of the analysis of the quality of strategic spreadsheets in use in our partner company, we were interested in evaluating the approach we developed for analysing spreadsheet quality. As spreadsheet users are application experts, we do not want to put too heavy a burden on them by requiring to switch from their "culture" as application experts to the "culture" of professional software developers. Nevertheless, they act as professional software developers when writing and maintaining long-living spreadsheets.

To assess our auditing technique's effectiveness, one has to recognise that there are two dimensions of freedom to be considered: The number of actual errors in the sheets available and the degree to which such errors are identified, and the effort needed to find those errors.

Obviously, testing and other conventional forms of software quality assurance can never demonstrate that the artefact analysed is faultless. Testing can only show that it finds faults. In our case, the auditor first analysed the sheets on the value dimension and found extremely few errors. This can be taken as indicator of the general high quality of the sheets. The ones he caught, though, can be taken as evidence for his careful checking and sufficiently mastering the application area. Looking on the model dimension, however, he found an overall error rate of 3,03 %. This not only meets our expectations, it is also consistent with results from other studies [Panko, 2000], [Panko, 1997b].

The second aspect is efficiency. The auditor who was no domain expert, stayed for 4 months at the company and actually spent 10 weeks on the audit. Hence, the examination of totally 60.446 cells was done in ten weeks by somebody who is not a domain expert. Of course, the errors identified were discussed with the sheets' authors, and documentation work had to be done. This gives an average inspection rate of 1208 cells per day.

Compared to other approaches (see [Panko, 1997]) this is rather high. Hence, we claim that the approach is worthwhile to follow at least for those portions of sheets, where high regularity is to be assumed and that complexity of size is well addressed. The structural complexity, however, is still an issue warranting further investigations.

## 6 DISCUSSION

The main task of the audit was twofold. On the face value, our industry partner wanted to have the companies spreadsheet audited (To be honest: Before we started, they were convinced that we would not find anything!). We, on the other hand wanted to assess the feasibility and effectiveness of the approach to audit spreadsheets on the basis of visualization by logical equivalence classes.

Concerning the first aspect, we might say that the quality of the company's spreadsheet was surprisingly good at first sight. The audit did not reveal spectacular wrong results. This might be due to the fact, that the spreadsheets are properly tested. However, they test only in the value domain and the correction on the value level made the spreadsheet model inconsistent. This bears the danger of spectacular errors to come up in future evolution steps. However, the audit still discovered 241 quantitative errors in the spreadsheets.

The company's representatives were very concerned of the audit's result. They stated that better spreadsheet development practices are going to be introduced. The representatives were also interested in guidelines to decide, whether a specific application should be realized by a spreadsheet or by a database application. One of the suggested improvements was better documentation and the application of systematic testing and auditing approaches.

The efficiency and performance of testing can be increased by use of a standardized auditing or testing methodology, as described in [Rothermel, 2000] or in [Ayalew, 2002]. The efficiency can be further increased by model visualization (see [Mittermeir, 2002]).

Insufficient documentation turned out to be the main cause of errors. Thus, we are currently working on guidelines for the documentation of spreadsheets. The lack of understanding due to missing documentation can even make some spreadsheets useless, if the maintainer leaves the company. Better understanding can be gained either by decreasing the overall complexity of the spreadsheet with design restrictions (see [Knight, 2000], [Isakowitz, 1995], [Wilde, 1993]), by giving a more comprehensive description of the spreadsheet (see [Paine, 1997], [Stadelmann, 1993]) or by visualizing the logical structure (see [Sajaniemi, 2000], [Chan, 2000], [Mittermeir, 2002]).

## 7 FUTURE WORK

Currently we are improving our auditing tool by a seamless integration of the dependency viewer. We aim to place it into one of the next releases of the open-source spreadsheet system *Gnumeric*. Our plans to integrate the toolkit with *Excel* are currently stalled, as we do not have access to the excel-formula-parser, while comparing parse-trees is a main issue of our toolkit.

We aim to support the auditing of large spreadsheets by adding further abstraction mechanisms to our approach. Among other things, we suggest to find groups of similar cells with similar neighbours and group them into semantic classes. Again, these semantic classes can be used for spotting irregularities in the spreadsheet.

# 8 CONCLUSION

This paper presents an auditing toolkit for assessing the correctness of large spreadsheets. The tool helps to identify irregularities in the spatial distribution of similar formulas. An assessment in an industrial context proved to be quite encouraging. It helped to analyse 78 spreadsheets, amongst them 62% contained errors. The cell error rate was 3.03 %. For the auditing itself, 4 person-months have been spent.

It turned out that the toolkit is suitable for auditing spreadsheets with large uniform or regular blocks by reducing the complexity of size. The auditors attention is focused to those cells were the regularity of formula occurrences is interrupted.

The main error sources we identified were the lack of documentation, maintenance and error corrections that were not consistent with the spreadsheet's internal logic. Thus, further ways for supporting spreadsheet comprehension are called for.